\documentclass[conference]{IEEEtran}
\IEEEoverridecommandlockouts
\usepackage{cite}
\usepackage{amsmath,amssymb,amsfonts}
\usepackage{algorithmic}
\usepackage{graphicx}
\usepackage{textcomp}
\usepackage{xcolor}
\usepackage{booktabs, tabularx}
\usepackage[referable]{threeparttablex}
\usepackage{adjustbox}
\usepackage{stfloats}% for placement table on the same page

\usepackage{array}
\usepackage{rotating}
\usepackage{multirow}
\usepackage{xcolor} 
\usepackage{soul}

\def\BibTeX{{\rm B\kern\checkmark.05em{\sc i\kern-.025em b}\kern-.08em
    T\kern-.1667em\lower.7ex\hbox{E}\kern-.125emX}}
\begin{document}

\title{AI Literacy for All: Adjustable Interdisciplinary Socio-technical Curriculum\\
% {\footnotesize \textsuperscriptNote: Sub\checkmarktitles are not captured in Xplore and
% should not be used}
% \thanks{Identify applicable funding agency here. If none, delete this.}
 }

\author{\IEEEauthorblockN{
\\Sri Yash Tadimalla\IEEEauthorrefmark{1},  
Mary Lou Maher\IEEEauthorrefmark{1},
}
    \\ 
    \IEEEauthorblockA{\IEEEauthorrefmark{1}University of North Carolina at Charlotte
    \\Charlotte, U.S.A.
    \\{stadimal, mmaher9} @uncc.edu}
    \\
    }

\maketitle
\begin{sloppypar}

\begin{abstract}
This research-to-practice paper presents a curriculum, "AI Literacy for All," to promote an interdisciplinary understanding of AI, its socio-technical implications, and its practical applications for all levels of education. With the rapid evolution of artificial intelligence (AI), there is a need for AI literacy that goes beyond the traditional AI education curriculum. AI literacy has been conceptualized in various ways, including public literacy, competency building for designers, conceptual understanding of AI concepts, and domain-specific upskilling. Most of these conceptualizations were established before the public release of Generative AI (Gen-AI) tools such as ChatGPT. AI education has focused on the principles and applications of AI through a technical lens that emphasizes the mastery of AI principles, the mathematical foundations underlying these technologies, and the programming and mathematical skills necessary to implement AI solutions. The non-technical component of AI literacy has often been limited to social and ethical implications, privacy and security issues, or the experience of interacting with AI. In AI Literacy for all, we emphasize a balanced curriculum that includes technical as well as non-technical learning outcomes to enable a conceptual understanding and critical evaluation of AI technologies in an interdisciplinary socio-technical context. 

The paper presents four pillars of AI literacy: understanding the scope and technical dimensions of AI, learning how to interact with Gen-AI in an informed and responsible way, the socio-technical issues of ethical and responsible AI, and the social and future implications of AI. While it is important to include all learning outcomes for AI education in a Computer Science major, the learning outcomes can be adjusted for other learning contexts, including, non-CS majors, high school summer camps, the adult workforce, and the public. This paper advocates for a shift in AI literacy education to offer a more interdisciplinary socio-technical approach as a pathway to broaden participation in AI. This approach not only broadens students' perspectives but also prepares them to think critically about integrating AI into their future professional and personal lives.

\end{abstract}


\section{Introduction}

The rapid evolution of artificial intelligence (AI) technology has ushered in transformative changes across numerous sectors, bringing new challenges and opportunities. As these technologies become increasingly embedded in everyday life, AI literacy is increasingly essential in today’s technology-driven world. According to a report by McKinsey (2020), approximately 70\% of businesses will adopt at least one form of AI technology by 2030, highlighting the growing integration of these systems into professional environments \cite{cam2019global}. This widespread adoption underscores the need for comprehensive AI literacy that extends beyond technical experts to the general population \cite{burgsteiner2016irobot}. AI literacy equips individuals not only with the skills to use AI tools effectively and safely but also with the ability to critically understand AI implications, including recognizing biases and potential ethical issues \cite{ng2021ai}. Furthermore, AI literacy is vital for workforce readiness in an economy increasingly reliant on AI technologies, ensuring individuals can adapt to new roles and job requirements. An informed public is essential for meaningful participation in debates and decision-making processes regarding AI governance and policy, advocating for AI developments that uphold public interest and ethical standards \cite{Firth-Butterfield_Toplic_Anthony_Reid_2022}. 

This paper proposes a significant shift from traditional AI education approaches towards an interdisciplinary socio-technical approach to an AI literacy curriculum that expands across the K-Grey lifelong learning model \cite{maher2024increasing}. By integrating insights from teaching AI concepts across various ages, disciplines, and current pathways into the field of AI, this approach aims to broaden participation in AI. It offers learners a more comprehensive view of how AI technologies are shaped and, in turn, shape our world, balanced with a perspective of personal agency. In advocating for the shift to the idea that "Everyone is an expert in their experience of AI," it is crucial to underscore the importance of preparing citizens not only as users and developers of AI technologies but also as informed contributors to the ongoing discourse around the ethical, social, and technical challenges posed by AI. This expanded perspective is a critical foundation for equipping all individuals with the ability to integrate AI into their professional and personal lives thoughtfully.

\section{Background}
The conceptualization of AI literacy is closely connected to the broader concept of digital literacy, which has been recognized as essential for all individuals navigating modern technology-driven environments \cite{celik2023exploring}. Just as digital literacy encompasses skills ranging from basic computer use to complex problem-solving in digital contexts, AI literacy must cover a spectrum from understanding simple AI functions to engaging with AI's broader implications for society. AI's prevalence in various sectors like finance, healthcare, and media underscores its broad utility and potential societal consequences, necessitating a comprehensive understanding and responsible application. 

AI literacy is not just an educational asset but a crucial component of informed citizenship and professional competence in the 21st century. It is important to note that the foundational frameworks for AI literacy were designed before the advent of advanced Gen-AI tools like ChatGPT, which have significantly altered the landscape of AI interaction for the general public \cite{touretzky2019envisioning}\cite{grover2024teaching}. These tools have made AI far more accessible to non-specialists, highlighting a critical gap in current educational paradigms. AI education has traditionally been approached primarily through a technical lens, focusing on the mastery of AI principles, the mathematical foundations of these technologies, and the programming skills required for their implementation \cite{eaton2024artificial} \cite{southworth2023developing}. This conventional curriculum has served well in fostering competency among designers and developers and enabling domain-specific upskilling \cite{kumar2024computer}. Enhancing public understanding of basic AI often focuses on the non-technical aspects of AI, including discussions on ethical implications, privacy, and security. This separation of technical vs non-technical AI literacy is insufficient in today’s socio-technical environment, where the implications of AI extend far beyond the code and into the fabric of societal norms and individual behaviors \cite{cultureperception2024}. 

Effective AI literacy now requires a more holistic approach, one that equally emphasizes the conceptual understanding and critical evaluation of AI technologies within an interdisciplinary social framework \cite{tadimalla2024implications}. Recent updates to AI education reflect these changes, emphasizing neural networks, practical AI applications, and the ethical, fairness, and transparency issues surrounding AI. Additionally, there is a concerted effort to enhance AI literacy and critical thinking across all areas of computer science education, linking AI closely with data science and maintaining a balanced view of symbolic and deep learning AI methods. Such a balance is crucial not only for technical competency but also for fostering an informed critical appreciation of the socio-technical dynamics at play. The types of AI literacy vary significantly depending on the target audience. For example, K-12 students are introduced to the basic concepts of AI and its everyday applications, fostering an early awareness that can influence future educational and career choices. Undergraduate programs, particularly in computer science, often focus on deep technical training, preparing the next generation of AI developers and researchers \cite{maher2024increasing}. However, AI literacy for non-CS majors is still nascent, emphasizing the need for a curriculum that integrates AI understanding across disciplines such as humanities, social sciences, and business \cite{laato2020propagating}. For the general public, AI literacy encompasses both career reskilling—vital for those in industries transformed by AI technologies—and general education aimed at making informed decisions about AI usage in personal life. This broad approach ensures that AI literacy extends from the classroom into lifelong learning pathways, crucial for adapting to continuous technological advancements.

A literature review of AI literacy reveals a diverse array of conceptual frameworks and pedagogical strategies aimed at enhancing the understanding and application of artificial intelligence across various educational settings \cite{ng2021ai} \cite{kong2024developing} \cite{su2023artificial}. These AI literacy reviews often provide a comprehensive understanding of how AI literacy is being defined, taught, and assessed, highlighting the evolving nature of this field \cite{yue2022pedagogical} \cite{schuller2022data} \cite{druga2019inclusive}. AI literacy frameworks often include four key aspects: knowing and understanding AI, using and applying AI, evaluating and creating with AI, and addressing ethical issues related to AI \cite{ng2021conceptualizing} \cite{touretzky2019envisioning} \cite{long2020ai}. These frameworks are based on the adaptation of classic literacies and aim to provide a consolidated definition and pedagogical direction for AI literacy education, particularly focusing on ethical concerns and competency development.  Some researchers argue for integrating AI literacy into the broader framework of technological literacy, viewing it as a multi-literacy that encompasses technical skills, technological scientific knowledge, and socio-ethical understanding \cite{stolpe2024artificial}\cite{kreinsen2023towards}. Some studies critically analyzed various components of AI literacy found in the literature and suggest that while technical skills are essential, greater emphasis should be placed on the socio-ethical aspects and the role of humans in AI, reflecting the need for a holistic approach to AI education \cite{stolpe2024artificial} \cite{casal2023ai} \cite{alvarez2022socially}. Evaluation of AI literacy for university students with diverse backgrounds highlights the potential of such courses to foster a conceptual understanding of AI without requiring prior programming knowledge \cite{kong2021evaluation}: future educational initiatives are recommended to consider AI as a fundamental skill within the 21st-century literacy framework, alongside reading, writing, arithmetic, and digital skills. Inspired by Bloom's taxonomy, the competencies for AI literacy should encompass basic cognitive abilities to know and understand AI, as well as advanced skills to use, apply, evaluate, and create AI solutions. Having various types of approaches is crucial for preparing individuals to meet future technological challenges and engage responsibly with AI technologies \cite{}.

 AI literacy can be categorized into several types, each focusing on different aspects of understanding and interacting with artificial intelligence technologies. These categorizations can help in tailoring educational programs to diverse audiences, from K-12 students, and professionals in various fields to members of society. Below are some types of AI literacy based on the lens with which researchers talk about AI literacy/education:\\
\textbf{Technical AI Literacy/Education:} This encompasses the foundational skills necessary to understand and develop AI systems, including programming, machine learning algorithms, and data science models. This type of AI literacy is emphasized in computer science education and professional development for AI specialists \cite{ng2021conceptualizing} \cite{kreinsen2023towards}.\\
\textbf{Gen-AI Literacy:} With the rise of Gen-AI technologies like ChatGPT, Bing, and SORA, etc this literacy type focuses on understanding and interacting with AI systems that can generate text, images, audio, video or other media. It involves knowledge of how these models are trained and their potential for both benign use and misuse \cite{chen2023generative}.\\
\textbf{General AI Literacy:} Focused on the everyday use and interaction with AI systems, this literacy type helps non-specialists understand how AI applications work in daily life, such as in smartphones, home assistants, and online services. It equips users to critically assess AI tools and their outputs in their environments \cite{long2020ai} \cite{kong2024developing}.\\
\textbf{Ethical and Social AI Literacy:} This addresses the socio-ethical implications of AI, such as privacy, security, fairness, and transparency. This type of literacy involves understanding the impact of AI technologies on society and considering ethical dilemmas associated with AI deployment and development \cite{zhang2023integrating}\cite{aiidentity2024}.\\
\textbf{Cognitive and Meta-cognitive AI Literacy:} This emphasizes the cognitive skills needed to interact with and adapt to advanced AI systems, including problem-solving and decision-making in contexts influenced by AI. This type of literacy also covers meta-cognitive skills that enable individuals to reflect on their learning processes and understand how AI can enhance their cognitive abilities. This prepares AI users for a future where immersive experiences and AI-driven projections across virtual and physical worlds could lead to significant shifts in how we perceive reality and engage with technology\cite{cao2023navigating} \cite{sokolowska2023impact} \cite{herath2024navigating}.

Along with developing an AI curriculum, we need further research on empirical and interventional study designs for various AI literacy interventions as underscored by the predominance of exploratory research in the field. This shift towards more rigorous research methodologies aims to address the quality of AI literacy assessments and develop definitive frameworks for educational practice. AI literacy assessment is a crucial component of this educational framework as it gauges the extent to which students have absorbed AI concepts and can apply them in various contexts. Effective assessment strategies ensure that AI literacy encompasses more than just technical competence; they also measure understanding of ethical implications, ability to engage with AI critically, and application of AI in solving real-world problems \cite{kong2024developing}. Given the interdisciplinary nature of AI, these assessments often blend technical tasks with scenario-based evaluations where students must navigate socio-technical challenges. Efforts to develop and validate assessment tools are pivotal \cite{laupichler2023development}. These exploratory analyses help identify the underlying factors of AI competence, contributing to the reliable and valid assessment of AI literacy \cite{knoth2024developing}. This approach underscores a trend towards more empirical and robust assessment methodologies that not only test knowledge but also the application and ethical considerations of AI, reflecting a comprehensive view of what it means to be AI literate in today's rapidly advancing technological landscape. 

In summary, the literature on AI literacy points to a dynamic field that is rapidly evolving to meet the needs of a technologically advanced society. The studies reviewed here collectively emphasize the importance of a multidimensional approach to AI literacy that incorporates not only technical skills evaluation but also assessing ethical, social, and cognitive competencies. This holistic approach is crucial for preparing educated citizens to effectively engage with AI technologies in their personal and professional lives, ensuring they are not only users but also informed, ethical decision-makers in an AI-driven world.

\section{ Socio-Technical AI Literacy Curriculum Learning Objectives}

With the proliferation of Gen-AI applications and the disruption caused by the disproportionate access and usage in learning environments, our curriculum is framed by the following questions that we believe every individual in our society should be competent to address across ages, disciplines, sectors, and demographics: What is AI and how does it work?
What are the ethical issues in AI? How should AI be used by individuals? How will AI impact our lives? 

The methodology for building this curriculum began with a literature review on AI literacy, focusing on identifying existing frameworks that outline key concepts taught by educators across the spectrum of technical and socio-technical approaches in AI literacy and AI education courses across the various types of emerging AI literacies mentioned above \cite{ng2021conceptualizing}\cite{touretzky2019envisioning}\cite{eaton2024artificial}  \cite{southworth2023developing} \cite{long2020ai} \cite{laupichler2022artificial}. The insights from the literature review (as discussed in the background section) were further supplemented by applying the Delphi method \cite{linstone1975delphi} to presentations, observations, and discussions from an NSF-funded workshop "Increasing Diversity in Lifelong AI Education" organized and facilitated by the authors \cite{maher2024increasing}, and the 2024 AAAI Spring Symposium on "Increasing Diversity in AI Education and Research" co-organized by the second co-author \cite{aaaiss2024}. These 3-day workshops on AI education and literacy brought together researchers, policy experts, and educators to respond to the increasing concerns and opportunities raised by recent AI developments, and to discuss directions for lifelong AI education. Our curriculum was then refined based on experiences from teaching this content in various settings, including a 2-credit general education course, 3-credit CS-major undergraduate and graduate courses, and a summer camp for middle and high school students. 

We present 4 Pillars of AI Literacy, shown in Table \ref{tab:CS2023}, based on a synthesis of the literature and our experience teaching AI Literacy. We use the taxonomy developed in the CS2023 curriculum to present our curriculum for easier integration into current efforts on curriculum alignment by ACM/IEEE-CS/AAAI \cite{eaton2024artificial}. These four pillars of socio-technical AI literacy are a foundation for an AI literacy curriculum that can be adapted for learning objectives providing a broad but defined vision for AI literacy efforts similar to the work by Touretzky (2019) for AI education \cite{touretzky2019envisioning}. The 4 pillars of AI Literacy are:  
 \begin{enumerate}
     \item Understanding the scope and technical dimensions of AI.
     \item Learning how to interact with Gen-AI in an informed and responsible way.
     \item Critically Reviewing the Issues of Ethical and Socially Responsible AI in Learning/Work Environments.  
     \item Social and Future Implications of AI.
 \end{enumerate}
 
The learning objectives for this interdisciplinary curriculum focus on providing students with a comprehensive understanding of AI's scope, technical dimensions, and ethical implications. The curriculum begins with an overview of AI, including its definitions, history, and current developments, particularly emphasizing recent advancements in deep learning that have led to applications like ChatGPT. Students will explore AI's fundamental concepts, including machine learning, natural language processing, computer vision, and robotics, gaining insights into the diverse subfields within AI. As students engage with AI technologies, they will learn to critically evaluate the benefits and limitations of AI in educational or professional settings. This includes understanding how AI can enhance learning through personalized experiences, data-driven insights, and increased accessibility, while also recognizing the potential risks, such as bias, privacy concerns, and the need for human oversight. The curriculum also emphasizes ethical and socially responsible AI, encouraging students to examine issues like algorithmic fairness, data security, and the broader societal impacts of AI on education and employment. Students will gain hands-on experience with AI tools, such as ChatGPT, to understand their capabilities and limitations in supporting learning. By critically reflecting on their own use of these tools, students will develop the skills to assess AI's role in their education and its potential implications in broader social contexts. Ultimately, the curriculum aims to equip students with the knowledge and critical thinking skills necessary to engage thoughtfully and responsibly with AI technologies in their educational, professional, and personal lives.
\begin{table}[h!]
\centering
\caption{ The 4 pillars of AI Literacy and topics presented using CS2023 Curricula taxonomy}
    \label{tab:CS2023}
    
\begin{tabular}{|>{\centering\arraybackslash}p{0.3\linewidth}|>{\centering\arraybackslash}p{0.5\linewidth}|>{\centering\arraybackslash}p{0.07\linewidth}|}
\hline
Pillar (Knowledge Unit)& Topic & CS/KA units \\
\hline
\multirow{4}{=}{Understanding the scope and technical dimensions of AI} & Introduction to AI and Machine Learning & 6
\\
\cline{2-3}
 & Representing knowledge in AI: symbolic and connectionist & 6
\\
\cline{2-3}
 & Search engines, generative systems, and retrieval augmented generation & 6
\\
\cline{2-3}
 & How do Large Language Models work? & 6
\\
\hline
\multirow{3}{=}{Learning how to interact with Generative AI in an informed and responsible way} & Interacting with Large Language Models & 6
\\
\cline{2-3}
 & Academic/Professional  Integrity, Authorship and Ownership & 6
\\
\cline{2-3}
 & Prompt engineering for learning & 6
\\
\hline
\multirow{5}{=}{Critically Reviewing the Issues of Ethical and Socially Responsible AI in Learning Environments}& Responsible use of AI & 5
\\
\cline{2-3}
 & Security, privacy, and ethical issues in AI & 6
\\
\cline{2-3}
 & Case Studies & 1\\
 & &\\
\hline
\multirow{4}{=}{Social and future implications of AI} & Public Perception of AI & 6
\\
\cline{2-3}
 & Generative AI and the Future of Work & 6
\\
\cline{2-3}
 & AI and Policy, Case Study of Accessibility & 6
\\
\cline{2-3}
 & AI for Good, Sustainability and Development & 4
\\
\hline
\multicolumn{2}{|c|}{Total} &  76\\
\hline
\end{tabular}

\end{table}

\begin{table*}[h!]
\centering

    \begin{threeparttable}
\label{tab:pillar1}
\caption{Topics and Learning Outcomes across Educational Levels, Pillar 1: Understanding the scope and technical dimensions of AI}
\small
\begin{tabular}{|>{\raggedright\arraybackslash}p{0.1\linewidth}|>{\centering\arraybackslash}p{0.3\linewidth}|>{\centering\arraybackslash}p{0.06\linewidth}|>{\centering\arraybackslash}p{0.06\linewidth}|>{\centering\arraybackslash}p{0.06\linewidth}|>{\centering\arraybackslash}p{0.06\linewidth}|>{\centering\arraybackslash}p{0.06\linewidth}|>{\centering\arraybackslash}p{0.06\linewidth}|>{\centering\arraybackslash}p{0.07\linewidth}|}
    \hline
    & & Non-CS Major*& \multicolumn{2}{c|}{CS Major*} & \multicolumn{2}{c|}{K-12} & \multicolumn{2}{c|}{Public} \\
    \hline
         Topic& Learning Outcome (Learning Outcome (CS/KA unit)) & Undergrad & Undergrad & Graduate & Middle School& High School*& Reskilling & Civic Education \\
    \hline
    \multirow{6}{=}
{Introduction to AI and Machine Learning}& What is AI  & 1 & 1 & 1 & 1 & 1 & 1 & 1 \\
        \cline{2-9}
        & The field and subfields of AI & 1 & 1 & 1 & 1 & 1 & 1 & 1 \\
        \cline{2-9}
        & Applications of AI & 1 & 1 & 1 & 1 & 1 & 1 & 1 \\
        \cline{2-9}
        & Machine Learning and Types & 1 & 1 & 1 & 1 & 1 & 1 & \\
        \cline{2-9}
        & History of AI & 1 & 1 & 1 & 1 & 1 & & 1 \\
        \cline{2-9}
        & AI in Education vs AI Education vs AI Literacy & 1 & 1 & 1 & 1 & 1 & & \\
    \hline 
    \multirow{6}{=}{Representing knowledge in AI: symbolic and connectionist} 
        & Symbolic AI & 1 & 1 & 1 &&& 1 &  \\
        \cline{2-9}
        & Connectionist AI & 1 & 1 & 1 &&& 1 &  \\
        \cline{2-9}
        & Neural Networks & 1 & 1 & 1 &&& 1 &  \\
        \cline{2-9}
        & Symbolic AI in Cognitive Architectures  &  & 1 & 1 &&&  &  \\
        \cline{2-9}
        & Neural Networks applications  &  & 1 & 1 &&&  &  \\
        \cline{2-9}
        & Deep Learning  & 1 & 1 & 1 &&& 1 &  \\
    \hline
    \multirow{6}{=}{Search engines, generative systems, and 
 RAG models }& Search Engines, How they work  & 1 & 1 & 1 && 1 & 1 &  \\
        \cline{2-9}
        & Generative AI Models and types & 1 & 1 & 1 && 1 & 1 &  1\\
        \cline{2-9}
        & Training Gen AI models  & 1 & 1 & 1 && 1 & 1 &  \\
        \cline{2-9}
        & Retrieval Augmented Generation (RAG)  & 1 & 1 & 1 &&& 1 &  \\
        \cline{2-9}
        & Training process for ChatGPT &  & 1 & 1 &&&&  \\
        \cline{2-9}
        & GPT Models 14 Journey  &  & 1 & 1 &&&&  \\
    \hline
    \multirow{6}{=}{How do Large Language Models work?} 
        & AI applications with LLMs & 1 & 1 & 1 & 1 & 1 & 1 &  1\\
        \cline{2-9}
        & Auto Encoders and Transformer models &  & 1 & 1 &&&&  \\
        \cline{2-9}
        & LLM Models Phases- Training and Generating & 1 & 1 & 1 &&& 1 &  \\
        \cline{2-9}
        & Pre-Training, Parameters, Features and Fine tuning  &  & 1 & 1 &&&&  \\
        \cline{2-9}
        & Natural Language Processing and Attention Mechanism &  & 1 & 1 &&&&  \\
        \cline{2-9}
        & Evaluation of Foundational AI Models  &  & 1 & 1 &&&&  \\
    \hline
    \multicolumn{2}{c}{Total Number of KA units}& 16& 24& 24& 7& 10& 14&6\\
    \hline
\end{tabular}
   \end{threeparttable}  
   
\end{table*}

\section{Adjustable AI Literacy For All}
Incorporating AI education across different levels of formal education and among various groups within the general public necessitates careful consideration of learning modalities, goals, and instructional methods to meet the diverse needs and backgrounds of learners \cite{maher2024increasing} \cite{su2023artificial} \cite{kandlhofer2016artificial}. Within the broad categories of K-12, higher education, and the general public, specific subgroups and populations require tailored approaches to AI education. Although AI topics are relevant across all educational levels, the depth and teaching modalities should vary to suit the unique needs of each group.

The curriculum presented in this paper is an additional Knowledge Area, as an alternative to the AI Knowledge Area, in the context of the CS2023 curriculum described by ACM/IEEE-CS/AAAI \cite{eaton2024artificial}. Tables II, III, IV, and V present the adjustable curriculum tailored to the goals and needs of AI literacy for learners in different educational contexts where the 4 pillars map as Knowledge Units, the learning outcomes as CS/KA core units as referred to in the CS2030 body of knowledge. The number of units for each topic reflects the inclusion and depth of coverage within each Pillar for specific educational contexts. While the curriculum is adjustable in terms of units, hours, and credits, we present this as a foundational AI Literacy curriculum. This is based on our experience teaching the content to non-CS majors, CS major undergraduates and graduates, and high school students under the K-12 category. The asterisk (*) in the tables serves as an indicator of the topics we have previously taught. Discussions in our development methodology inform the suggested units for middle school and the general public.

\textbf{In the K-12 or Elementry-Highschool education,} AI education starts with structured modalities, focusing on basic definitions and abstract concepts of AI technology. Elementary students are introduced to fundamental ideas, while middle and high school students progress to technical foundations and socio-technical concepts with increasing complexity. Activities are designed to be interactive and engaging, such as using block-based coding tools like Scratch to integrate AI modules, allowing students to see immediate outcomes of simple AI functionalities. Hands-on projects, like robotics or simple neural network simulations, help solidify understanding, followed by reflections on how AI impacts daily life \cite{druga2019inclusive}. Critiques at this stage focus on fairness and bias at a fundamental level.

\textbf{In higher education,} AI literacy goals differ markedly between CS majors and non-CS majors. CS majors are expected to demonstrate mastery and apply technical components of AI, primarily focusing on Pillars 1 and 2. In contrast, non-CS majors engage more with Pillars 3 and 4, emphasizing the ethical and social implications of AI. At the undergraduate level, students also focus on Academic/Professional  Integrity more than those in K-12 or general public programs. Graduate education emphasizes research and specialized applications of AI literacy, with a comprehensive approach to research papers, literature reviews, and community projects, varying by major and discipline. Modalities evolve to include deeper technical readings, complex lab activities, and domain-specific applications. For CS majors, activities might involve advanced programming and AI system problem-solving, while non-CS majors focus on applying AI tools within fields like healthcare or economics. Reflections encourage students to consider the ethical implications and societal impacts of their work.

\textbf{For the adult workforce,} the focus shifts towards continuous education and professional development, integrating AI training relevant to specific job roles. Readings are centered on case studies illustrating AI applications in business or industry settings. Workshop-oriented activities, often in online formats, emphasize the practical use of AI tools to optimize workflows or automate tasks. Reflections involve assessing the return on investment of AI integration and considering ethical implications relevant to their sectors \cite{laupichler2022artificial}.

\textbf{For the general public,} AI education takes on a more varied approach, including community workshops, public lectures, and online modules designed to cater to diverse groups. Introductory readings provide an overview of AI, highlighting its potential benefits and risks. Interactive sessions using AI in everyday applications, such as virtual assistants or recommendation systems, make the technology relatable. Public critiques focus on understanding AI news articles and debunking common AI myths, fostering critical media literacy. Reflections promote community discussions on the personal and societal impacts of AI, encouraging informed public discourse on AI policy and development.

Across all levels, the combination of readings, activities, labs, and reflections is structured to not only impart knowledge but also build a deeper understanding of AI’s capabilities and limitations, ethical considerations, and potential societal impact. The progression of these modalities ensures that each educational phase builds on the previous one, gradually increasing in complexity and application according to the learners' age, background, and professional needs. For the general public, the approach is designed for broad engagement and awareness, suitable for varying levels of prior knowledge and interest. 

\section{ Socio-Technical AI Literacy Learning Outcomes }
 
The course content is framed by the following topics to achieve the learning objectives. 
Topics: 

\textbf{Introduction to AI and Machine Learning}
The curriculum begins with a foundational introduction to Artificial Intelligence (AI) and Machine Learning, establishing a clear understanding of what AI entails. This section explores the expansive field and its subfields, including machine learning, natural language processing, robotics, and computer vision. The applications of AI are vast, spanning from autonomous vehicles and healthcare diagnostics to personal assistants like Siri and Alexa. A thorough examination of machine learning and its various types, such as supervised, unsupervised, and reinforcement learning, is provided. This topic area also explores the history of AI, charting its evolution from early rule-based representations to contemporary deep learning models. The role of AI in education is dissected into three distinct areas: AI in Education, AI Education, and AI Literacy, highlighting the transformative impact of AI on personalized learning, administrative automation, and educational content enhancement.

\textbf{Representing Knowledge in AI: Symbolic and Connectionist}
The curriculum transitions into the representation of knowledge in AI, focusing on both symbolic and connectionist approaches. Symbolic AI, or good old-fashioned AI (GOFAI), is explored in terms of its use of explicit rules and symbols to represent knowledge and logic, often employed in expert systems and cognitive architectures. In contrast, connectionist AI relies on neural networks that mimic the brain's neuron structures. Detailed coverage of neural networks and deep learning models is provided, emphasizing their role in revolutionizing AI through tasks like image and speech recognition. This section also discusses the integration of symbolic AI in cognitive architectures for complex reasoning and problem-solving, alongside the advantages of connectionist AI in pattern recognition and data-driven tasks.

\textbf{Search Engines, Generative Systems, and Retrieval Augmented Generation}
The curriculum examines search engines, generative systems, and Retrieval Augmented Generation (RAG). Students are introduced to the workings of search engines, including their indexing and retrieval processes based on user queries and relevance algorithms. Gen-AI models, such as the Generative Pre-trained Transformer (GPT), are discussed, with a focus on their ability to create new content by learning patterns from extensive datasets. The training process for these generative models is dissected, highlighting the importance of large data sets and parameter tuning. RAG combines generative models with traditional search engines, retrieving relevant documents to inform content generation, and enhancing the accuracy and relevance of the generated content, as seen in applications like ChatGPT.

\textbf{How Do Large Language Models Work?}
This topic area focuses on the mechanics of Large Language Models (LLMs), such as GPT-3 and GPT-4. These models use deep learning techniques to process and generate human-like text, leveraging transformer architectures and self-attention mechanisms to understand word relationships and context. The curriculum covers the training phases of LLMs, from pre-training on vast text corpora to fine-tuning for specific tasks. Key concepts such as pre-training, parameters, features, and fine-tuning are elucidated. The evaluation of foundational AI models is discussed, emphasizing their application in various AI domains like natural language processing and attention mechanisms.

\begin{table*}[h!]
\centering

    \begin{threeparttable}
\label{tab:pillar2}
\caption{Topics and Learning Outcomes across Educational Levels, Pillar 2:  Learning how to interact with Gen-AI in an informed and responsible way}
\small
\begin{tabular}{|>{\raggedright\arraybackslash}p{0.1\linewidth}|>{\centering\arraybackslash}p{0.3\linewidth}|>{\centering\arraybackslash}p{0.06\linewidth}|>{\centering\arraybackslash}p{0.06\linewidth}|>{\centering\arraybackslash}p{0.06\linewidth}|>{\centering\arraybackslash}p{0.06\linewidth}|>{\centering\arraybackslash}p{0.06\linewidth}|>{\centering\arraybackslash}p{0.06\linewidth}|>{\centering\arraybackslash}p{0.07\linewidth}|}
    \hline
    & & Non-CS Major*& \multicolumn{2}{c|}{CS Major*} & \multicolumn{2}{c|}{K-12} & \multicolumn{2}{c|}{Public} \\
    \hline
         Topic& Learning Outcome (CS/KA unit)& Undergrad & Undergrad & Graduate & Middle School & High School*& Reskilling & Civic Education \\
    \hline
    \multirow{6}{=}
{Interacting with Large Language Models}& Closed Source LLMs& & 1& 1& 1& 1& 1& 
\\
        \cline{2-9}
        & Open Source LLMs& & 1& 1& 1& 1& 1& 
\\
        \cline{2-9}
        & Gen AI applications & & 1& 1& 1& 1& & 
\\
        \cline{2-9}
        & Best Practices for interacting with LLMs& 1& 1& 1& 1& 1& 1& 
1\\
        \cline{2-9}
        & Prompts and responses & 1& 1& 1& 1& 1& 1& 
1\\
        \cline{2-9}
        & AI and writing & 1& 1& 1& 1& 1& 1& 
1\\
    \hline
    \multirow{6}{=}{Academic/ Professional  Integrity, Authorship and Ownership }& Why is Academic / Professional  Integrity Important?& 1& 1& 1&1&1& &  
\\
        \cline{2-9}
        & Code of Student Conduct& 1& 1& 1&&1& &  
\\
        \cline{2-9}
        & Types of Academic Misconduct& 1& 1& 1&&1& &  
\\
        \cline{2-9}
        & Reasons to avoid Misconduct&  1& 1& 1&&1&  &  
\\
        \cline{2-9}
        & Understanding Misconduct Sanctions and Policies&  1& 1& 1&&&  &  
1\\
        \cline{2-9}
        & Strategies to avoid Misconduct& 1& 1& 1&&& &  
\\
    \hline
    \multirow{6}{=}{Prompt engineering for learning}& Using AI for Learning Interventions& 1& 1& 1&1& 1& 1&  
\\
        \cline{2-9}
        & Using Search, Generative AI, and Retrieval Augmented Generation to study& & 1& 1&1& 1& &  
1\\
        \cline{2-9}
        & Responsible use of Gen AI for testing& & 1& 1&1& 1& &  
\\
        \cline{2-9}
        & Structure of a Prompt & 1& 1& 1&1&1& 1&  1
\\
        \cline{2-9}
        & Coding/Programming with AI tools &  1& 1& 1&&&1&  
\\
        \cline{2-9}
        & The End of Programming &  & 1& 1&&&&  \\
         \hline
     \multicolumn{2}{c}{Total Number of KA units}& 12& 18& 18& 11& 14& 8&6\\
     \hline
\end{tabular}
   \end{threeparttable}  
   
\end{table*}
\textbf{Interacting with Large Language Models}
In this section, the curriculum addresses the interaction with LLMs, differentiating between closed-source and open-source models. Closed-source LLMs, typically proprietary with sophisticated capabilities, are compared with open-source models, which offer transparency and community-driven enhancements. Best practices for interacting with LLMs are outlined, including the formulation of effective prompts, setting response parameters, and considering ethical implications. The curriculum emphasizes the practical applications of LLMs in writing and creative tasks, advocating for responsible use to maintain authenticity and originality.

\textbf{Academic/Professional  Integrity, Authorship, and Ownership}
The importance of Academic/Professional  Integrity is a pivotal theme, highlighting the necessity of ethical standards in research, writing, and assessments. The curriculum details the Code of Student Conduct and identifies behaviors constituting academic misconduct, such as plagiarism and cheating. It underscores the significance of understanding misconduct policies and sanctions, and provides strategies to avoid academic misconduct, fostering a culture of honesty and respect within academic communities.

\textbf{Prompt Engineering for Learning}
Prompt engineering is explored as a crucial technique for guiding AI responses in educational settings. This section demonstrates how AI can enhance learning interventions by providing personalized feedback and resources. It discusses the use of search engines, Gen-AI, and RAG for research and problem-solving. Responsible use of Gen-AI in testing is emphasized to ensure fairness and avoid dependency. Students learn to structure prompts effectively to elicit relevant AI responses, enhancing their learning experience and improving educational outcomes.

\begin{table*}[h!]
\centering

\small
    \begin{threeparttable}
\label{tab:pillar3}
\caption{Topics and Learning Outcomes across Educational Levels, Pillar 3: Critically Reviewing the Issues of Ethical and Socially Responsible AI in Learning Environments}
\begin{tabular}{|>{\raggedright\arraybackslash}p{0.1\linewidth}|>{\centering\arraybackslash}p{0.3\linewidth}|>{\centering\arraybackslash}p{0.06\linewidth}|>{\centering\arraybackslash}p{0.06\linewidth}|>{\centering\arraybackslash}p{0.06\linewidth}|>{\centering\arraybackslash}p{0.06\linewidth}|>{\centering\arraybackslash}p{0.06\linewidth}|>{\centering\arraybackslash}p{0.06\linewidth}|>{\centering\arraybackslash}p{0.07\linewidth}|}
    \hline
    & & Non-CS Major*& \multicolumn{2}{c|}{CS Major*} & \multicolumn{2}{c|}{K-12} & \multicolumn{2}{c|}{Public} \\
    \hline
         Topic& Learning Outcome (CS/KA unit)& Undergrad & Undergrad & Graduate & Middle School & High School*& Reskilling & Civic Education \\
    \hline
    \multirow{6}{=}
{Responsible use of AI}& Responsible Use of AI& 1& 1& 1&& 1& 1& 

1\\
        \cline{2-9}
        & Responsible Use of AI in  Education& & 1& 1&& 1& 1& 

\\
        \cline{2-9}
        & AI Compliance& & 1& 1&&& 1& 

\\
        \cline{2-9}
        & Risks of Over-Reliance on Generative AI
& & 1& 1&&& 1& 
\\
        \cline{2-9}
        & Case Studies on Responsible use of AI in Education& 1& 1& 1& 1& 1& 1& 

\\
    \hline
    \multirow{6}{=}{Security, privacy, and ethical issues in AI }& Cyber Security and AI & 1& 1& 1&1&1& 1&  
1
\\
        \cline{2-9}
        & 
Ethics and AI & 1& 1& 1&1&1& 1&  
1
\\
        \cline{2-9}
        & Co-Creative AI: Human AI collaboration & 1& 1& 1&1&1& 1&  
1
\\
        \cline{2-9}
        & 
Explainable/ Secure/ Ethical/ Human-Centered AI&  & 1& 1&1&1&  1&  

1
\\
        \cline{2-9}
        & AI Security and Privacy  vs Secure  AI
&  & 1& 1&&&  1&

\\
        \cline{2-9}
        & 
Critiquing and evaluating AI & & 1& 1&1&1& 1&  

\\
        \cline{2-9}
        & Case Studies &  1& 1& 1&1&1&1&  1\\
         \hline
         
    \multicolumn{2}{c}{Total Number of KA units}& 6& 12& 12& 7& 9& 12&6\\
     \hline
\end{tabular}
   \end{threeparttable}  
   
\end{table*}
\begin{table*}[h!]
\centering

\small
    \begin{threeparttable}
\label{tab:pillar4}
\caption{Topics and Learning Outcomes across Educational Levels,   Pillar 4: Social and future implications of AI }
\begin{tabular}{|>{\raggedright\arraybackslash}p{0.1\linewidth}|>{\centering\arraybackslash}p{0.3\linewidth}|>{\centering\arraybackslash}p{0.06\linewidth}|>{\centering\arraybackslash}p{0.06\linewidth}|>{\centering\arraybackslash}p{0.06\linewidth}|>{\centering\arraybackslash}p{0.06\linewidth}|>{\centering\arraybackslash}p{0.06\linewidth}|>{\centering\arraybackslash}p{0.06\linewidth}|>{\centering\arraybackslash}p{0.07\linewidth}|}
    \hline
    & & Non-CS Major*& \multicolumn{2}{c|}{CS Major*} & \multicolumn{2}{c|}{K-12} & \multicolumn{2}{c|}{Public} \\
    \hline
         Topic& Learning Outcome (CS/KA unit)& Undergrad & Undergrad & Graduate & Middle School & High School*& Reskilling & Civic Education \\
    \hline
    \multirow{6}{=}
{Public Perception of AI}& Perception and Mental Models& & 1& 1&&&& 
\\
        \cline{2-9}
        & Public Perception of AI& 1& 1& 1&&&& 1
\\
        \cline{2-9}
        & Culture Trust and AI& 1& 1& 1&&&& 1
\\
        \cline{2-9}
        & Misconceptions of AI& 1& 1& 1&& 1& 1& 1
\\
        \cline{2-9}
        & AI and Media & 1& 1& 1& 1& 1&& 1
\\
        \cline{2-9}
        & AI ecosystem and Identity& 1& 1& 1&&1&1& 
1\\
    \hline
    \multirow{6}{=}{Generative AI and the Future of Work}& Shifts in AI and Technology Industry& 1& 1& 1&1&1& 1&  1
\\
        \cline{2-9}
        & Digital Economy and ICT & 1& 1& 1&&1& 1&  \\
        \cline{2-9}
        & Opportunities vs Disruptions& 1& 1& 1&&1& 1&  1
\\
        \cline{2-9}
        & Un-AI-able: Human Vs AI &  & 1& 1&1&1&  &  1
\\
        \cline{2-9}
        & Gen AI and Future of work &  1& 1& 1&1&1&  1&  1
\\
        \cline{2-9}
        & AI in different Sectors and Fields& & 1& 1&1&1& &  1
\\
    \hline
    \multirow{6}{=}{AI and Policy, Case Study of Accessibility }& Types of AI Policy and Institutions& & 1& 1&1&&&  
\\
        \cline{2-9}
        & Why it Matters & 1& 1& 1&&&&  
\\
        \cline{2-9}
        & Science Policy: Policy and Technology& & 1& 1&&&&  
\\
        \cline{2-9}
        & History of AI Policy & & 1& 1&&&&  1
\\
        \cline{2-9}
        & Actors in AI Policy &  1& 1& 1&1&1&&  1
\\
        \cline{2-9}
        & Executive Orders &  1& 1& 1&&&&  1
\\
    \hline
    \multirow{6}{=}{AI for Good, Sustainability and Development}& Sustainable Development Goals & 1& 1& 1& 1& 1& 1&  1
\\
        \cline{2-9}
        & Carbon footprint and Energy cost of Technology &  1& 1& 1&1&1&1&  1
\\
        \cline{2-9}
        & AI for Social Good & 1& 1& 1&1&1& 1&  1
\\
        \cline{2-9}
        & Case Studies on AI for Good &  & 1& 1&1&1&&  \\
     \hline        
     \multicolumn{2}{c}{Total Number of KA units}& 15& 22& 22& 11& 14& 10&17\\
     \hline
     
\end{tabular}
   \end{threeparttable}  
   
\end{table*}

\textbf{Responsible Use of AI}
The curriculum addresses the responsible use of AI, emphasizing ethical considerations and regulatory compliance. In educational contexts, responsible AI usage entails enhancing learning while safeguarding privacy and equality. The curriculum covers AI compliance, including data usage guidelines and algorithm transparency, and warns against the risks of over-reliance on Gen-AI. Case studies on responsible AI usage in education provide practical examples of best practices, underlining the importance of ethical AI deployment to maximize benefits and mitigate potential drawbacks.

\textbf{Security, Privacy, and Ethical Issues in AI}
Students are introduced to the critical issues of security, privacy, and ethics in AI. This section covers cybersecurity measures to protect AI systems from malicious attacks and privacy safeguards for data protection. Ethical AI practices, including fairness, transparency, and accountability, are discussed. The curriculum explores human-AI collaboration, or co-creative AI, which leverages the strengths of both humans and machines. Concepts of explainable AI are introduced to enhance trust and usability. The importance of evaluating AI systems for security, privacy, and ethical standards is emphasized.

\textbf{Public Perception of AI}
The curriculum explores how public perception of AI is influenced by cultural, social, and media factors. Understanding mental models of AI helps developers create more user-friendly and trusted systems. This section addresses common misconceptions about AI and the importance of building trust through transparent communication. It also considers AI's impact on identity and culture, advocating for responsible representation to avoid reinforcing biases and stereotypes. Engaging the public in AI discourse is highlighted as a means to foster informed opinions and acceptance.

\textbf{Gen-AI and the Future of Work}
The transformative impact of Gen-AI on industries and the future of work is a key topic. This section discusses how AI-driven technological shifts create new opportunities and disrupt traditional job roles. The evolving digital economy and the role of Information and Communication Technology (ICT) are examined. Strategies for balancing opportunities with potential disruptions are explored to ensure a sustainable workforce. The curriculum underscores the importance of human skills that AI cannot replicate, highlighting the unique capabilities of humans versus AI.

\textbf{AI and Policy, Case Study of Accessibility}
AI policy is explored in-depth, covering regulations and guidelines that govern AI development and deployment. This section discusses the significance of different types of policies and institutions involved in AI governance. The curriculum explains why AI policy matters, addressing ethical, social, and economic implications. Historical perspectives on AI policy are provided, alongside discussions of key actors, including governments, organizations, and industry leaders. Executive orders and legislative actions are examined, emphasizing their impact on AI accessibility and inclusivity.

\textbf{AI for Good, Sustainability, and Development}
The curriculum concludes with an exploration of AI's potential for good, focusing on sustainability and development. It discusses how AI can contribute to achieving Sustainable Development Goals (SDGs) by addressing global challenges like poverty, healthcare, and education. The environmental impact of AI, including its carbon footprint and energy costs, is critically examined. Case studies on AI for social good demonstrate practical applications that drive sustainable development and improve quality of life. The curriculum advocates for balancing technological advancements with environmental and social considerations to achieve a sustainable and equitable future.

Developing distinct educational goals for AI across various levels is crucial for broad participation and understanding. For K-12 students, the focus is on equitable access, foundational AI principles, and hands-on experiences, coupled with strong ethics and critical thinking components to prepare a diverse future workforce. Post-secondary goals aim to imbue CS and non-CS majors with deep AI competence and ethical frameworks, highlighting the relevance of AI across disciplines and preparing them for responsible AI use in their careers. For the adult workforce, education targets adapting to AI-driven work changes, emphasizing skills enhancement and ethical considerations to manage career transitions effectively. Public AI education strives to foster informed citizenship by enhancing understanding of AI technologies, mitigating biases, and promoting ethical use, ensuring that all community members can engage responsibly with AI and appreciate its societal benefits. This comprehensive curriculum provides a robust foundation for understanding AI and its multifaceted impacts, equipping students with the knowledge and skills to navigate and contribute to the evolving landscape of AI technology while addressing the needs of learners at all stages of education and career development.

\end{sloppypar}

\section{Conclusion}

In conclusion, this paper advocates for a fundamental shift in how AI literacy is conceptualized and delivered. By adopting a comprehensive interdisciplinary socio-technical approach, AI literacy can be transformed into a foundational element of education at all levels, from K-12 to lifelong learning. The four-pillar approach to AI Literacy focuses on:  Understanding the scope and technical dimensions of AI; Learning how to interact with Gen-AI in an informed and responsible way; Critically Reviewing the Issues of Ethical and Socially Responsible AI in Learning/Work Environments; and Social and Future Implications of AI. This approach ensures that all citizens, including students in Computer Science majors, are equipped to navigate and influence the AI-augmented landscapes of the future responsibly and ethically. Our approach is unique in presenting a comprehensive curriculum for AI Literacy that is adjustable to serve the specific goals in a broad range of learning contexts.

% comprehending AI's fundamental concepts, architectures, and technical underpinnings, enabling students to grasp how various AI technologies are designed and function. Teaching students the skills necessary to effectively and ethically engage with Gen-AI technologies, emphasizing the importance of responsible usage. Emphasizing the personal responsibility of students to understand and address the ethical and social implications of AI technologies, particularly within educational settings, fosters an environment of conscientious learning and application. Exploring the broader societal and future-oriented consequences of AI deployment 

% \section*{Acknowledgment}

% The preferred spelling of the word ``acknowledgment'' in America is without 
% an ``e'' after the ``g''. Avoid the stilted expression ``one of us (R. B. 
% G.) thanks $\ldots$''. Instead, try ``R. B. G. thanks$\ldots$''. Put sponsor 
% acknowledgments in the unnumbered footnote on the first page.

\bibliographystyle{IEEEtran}
\bibliography{FIE_CAPACITI}

\end{document}